\documentclass[12pt]{article}
\usepackage{amsmath,amssymb}
\usepackage[dvipdfmx]{graphicx}
\usepackage{bm}
\usepackage{color}
\usepackage{comment}

\def\mydate{20 August 2015}
\def\ignore#1{{}}

\makeatletter
\@addtoreset{equation}{section}
\makeatother

\newcommand{\siml}{%
\hspace{0.3em}\raisebox{0.4ex}{$<$}\hspace{-0.75em}\raisebox{-.7ex}{$\sim$}\hspace{0.3em}}

\newcommand{\beeq}{\begin{equation}}
\newcommand{\eneq}{\end{equation}}
\newcommand{\beqn}{\begin{eqnarray}}
\newcommand{\eeqn}{\end{eqnarray}}

\def\dd{\partial}
\def\la{\raise.16ex\hbox{$\langle$}\lower.16ex\hbox{}  }
\def\ra{\raise.16ex\hbox{$\rangle$}\lower.16ex\hbox{} }
\def\go{\rightarrow}

\def\onehalf{ \hbox{$\frac{1}{2}$} }

\def\onethird{ \hbox{$\frac{1}{3}$} }

\def\Tr{{\rm Tr \,}}

\def\eff{{\rm eff}}
\def\min{{\rm min}}

\def\cL{{\cal L}}
\def\cD{{\cal D}}

\def\EM{{\rm EM}}

\def\diag{{\rm diag ~}}

\def\KK{{\rm KK}}

\def\ep{\epsilon}
\def\psibar{ \psi \kern-.65em\raise.6em\hbox{$-$} }
\def\psibarl{ \psi \kern-.65em\raise.6em\hbox{$-$} \lower.6em\hbox{} }

\begin{document}

\thispagestyle{empty}

{\small \noindent \mydate    \hfill OU-HET 854}


\vspace{4.0cm}

\baselineskip=20pt plus 1pt minus 1pt

\begin{center}
{\LARGE \bf  Gauge-Higgs Grand Unification}

\end{center}

\vspace{.0cm}
\baselineskip=22pt plus 1pt minus 1pt

\begin{center}
{\bf
Yutaka Hosotani and Naoki Yamatsu
}

\vskip 5pt

{\small \it Department of Physics, 
Osaka University, 
Toyonaka, Osaka 560-0043, 
Japan} \\

\end{center}


\vskip 2.5cm
\baselineskip=18pt plus 1pt minus 1pt

\begin{abstract}
$SO(11)$ gauge-Higgs grand unification in the Randall-Sundrum warped space is
proposed.  Orbifold boundary conditions and one brane scalar field reduce  $SO(11)$
to the standard model symmetry,  which is further broken to $SU(3)_C \times U(1)_\EM$ 
by the Hosotani mechanism.  In a minimal model quarks and leptons are contained 
in a  multiplet in  {\bf 32} of $SO(11)$ in each generation. 
Proton decay is naturally suppressed  by a conserved fermion number.
\end{abstract}

\ignore{\qquad \small
}



\newpage

\baselineskip=20pt plus 1pt minus 1pt
\parskip=0pt

The discovery of the Higgs boson at LHC supports the current scenario of the 
unification of electromagnetic and weak forces.   The electroweak  (EW) gauge 
symmetry, $SU(2)_L \times U(1)_Y$, is spontaneously broken to $U(1)_\EM$ 
by the VEV (vacuum expectation value) of the Higgs scalar field.  
All experimental data so far are consistent with the standard model (SM) of 
electroweak and strong interactions.  Yet it is not clear whether the observed
Higgs boson is precisely what the SM predicts.  Detailed study of the interactions
among the Higgs boson and other SM particles  in the coming experiments is in desperate
necessity.

There remain uneasy features in the Higgs boson sector in the SM.  
 Unlike such gauge bosons as photon, $W$ boson, $Z$ boson and gluons, 
 whose dynamics is governed by the gauge principle, 
 the Higgs boson is an elementary scalar field for which there lacks an underlying 
 fundamental principle.
 The Higgs couplings of quarks and leptons as well as the Higgs self-couplings are not
 regulated by any principle.
 At the quantum level there arise huge corrections to the Higgs boson mass
 which has to be cancelled and tuned by hand to obtain the observed 125$\,$GeV mass.
One way to achieve natural stabilization of the Higgs boson mass against quantum corrections
is to invoke supersymmetry, and many investigations have been made along this line.
In this paper we focus on an alternative approach, the gauge-Higgs unification.\cite{YH1, Davies1, Hatanaka1998}

The Higgs boson is unified with gauge bosons in the gauge-Higgs unification, which is formulated 
as a gauge theory in five or more dimensions.  When the extra-dimensional space is not simply
connected, an Aharonov-Bohm (AB) phase in the extra-dimensional space plays the role of the 
Higgs boson, breaking a part of non-Abelian gauge symmetry.  The four-dimensional (4D) fluctuation
mode of the AB phase appears as a Higgs boson in four dimensions at low energies.  
Put in other words, the Higgs boson is a part 
of the extra-dimensional component of gauge potentials, whose dynamics is controlled by the 
gauge principle.  The gauge invariance guarantees the periodic nature of physics associated 
with the AB phase in the extra dimension which we denote as $\theta_H$.
 
The value of $\theta_H$ is determined dynamically, from the location of the global minimum 
of the effective potential $V_\eff (\theta_H)$.  At the classical (tree) level $V_\eff (\theta_H)$ is
completely flat, as $\theta_H$ is an AB phase yielding vanishing field strengths.  
At the quantum level $V_\eff (\theta_H)$ becomes nontrivial as the particle spectrum and 
their interactions depend on $\theta_H$.  It has been shown  that 
the $\theta_H$-dependent part of $V_\eff (\theta_H)$ is finite at the one loop level, free from
ultraviolet divergence even in five or more dimensions as a consequence of the gauge invariance.  
Nontrivial minimum $\theta_H^\min$ induces gauge symmetry breaking in general.
The mass of the corresponding 4D Higgs boson, proportional to the second derivative of 
$V_\eff (\theta_H)$ at the minimum, becomes finite irrespective of the cutoff scale in a theory, 
giving a way to solve the gauge hierarchy problem.  This mechanism of dynamical gauge
symmetry breaking is called as the Hosotani mechanism.

Gauge-Higgs unification models of  electroweak interactions have been 
constructed.\cite{ACP2005}--\cite{LHCsignalsDM}
The orbifold structure of the extra-dimensional space is vital to have chiral fermions, and
natural realization of dynamical EW symmetry breaking is achieved in the five-dimensional 
Randall-Sundrum (RS) warped spacetime.  The most promising is the $SO(5) \times U(1)_X$ 
gauge-Higgs unification in RS, which is  consistent with the observation at low energies
provided  its AB phase $\theta_H \siml 0.1$. 
The model accommodates the custodial symmetry, and gives almost the same couplings 
in the gauge sector as the SM.
It has been shown that 1-loop corrections to the Higgs boson decay to $\gamma \gamma$ 
due to running of an infinite number of Kaluza-Klein (KK) excitation modes of $W$ boson 
and top quark turn out finite and very small, being consistent with the present LHC data.\cite{FHHOS2013}
It predicts Kaluza-Klein excitations of $Z$ boson and photon as $Z'$ events with broad widths 
in the mass range $5\,$TeV to $8\,$TeV, and a dark matter candidate (dark fermion) of  a mass 
$2\,$TeV to $3\,$TeV,
and other signals such as anomalous Higgs couplings 
are predicted as well.\cite{LHCsignalsDM}-\cite{Maru2015}

With the gauge-Higgs EW unification model at hand, the next step is to incorporate
strong interactions to achieve gauge-Higgs grand unification.\cite{Kawamura2000}-\cite{Yamamoto2014}
There are models of gauge-Higgs grand unification in five dimensions with gauge group $SU(6)$,
which breaks down to $SU(3)_C \times SU(2)_L \times U(1)_Y \times U(1)_X$ by
the orbifold boundary condition on $S^1/Z_2$.  
Burdman and Nomura showed that the EW Higgs doublet emerges.  
Haba et al.\  and Lim and Maru  showed that dynamical 
EW symmetry is achieved with extra matter fields, though they yield exotic particles at low energies.
Kojima et al.\ have proposed an alternative model with $SU(5) \times SU(5)$ symmetry.
Grand unification in the composite Higgs scenario has been discussed by Frigerio et al.
Yamamoto has  attempted to dynamically derive orbifold boundary conditions 
in gauge-Higgs unification models. 

In this paper we propose a new model of gauge-Higgs grand unification in RS with gauge symmetry 
$SO(11)$ which carries over good features of $SO(5) \times U(1)_X$ gauge-Higgs EW unification.
We show that the  EW symmetry breaking is induced even in the pure gauge theory 
by the Hosotani mechanism, in sharp contrast to other models.
Quarks and leptons are implemented in a minimal set of fermion multiplets.
Proton decay is naturally suppressed by the conservation of a new fermion number.

The model is defined in the RS spacetime with  metric 
$ds^2 = e^{-2 \sigma(y)} \eta_{\mu\nu} dx^\mu dx^\nu + dy^2$, where
$\eta_{\mu\nu} = \diag (-1, 1, 1, 1)$, 
$\sigma (y) = \sigma(-y) = \sigma(y+ 2L)$, and $\sigma(y) = k y$ for $0 \le y \le L$.
$z_L = e^{kL} \gg 1$ is called the warp factor.
The AdS space with a cosmological constant $\Lambda = - 6k^2$ in $0 < y < L$ is 
sandwiched by the Planck brane  at $y=0$ and  the TeV brane at $y=L$.

The $SO(11)$ gauge potential, $A_M$, expressed as 11 by 11 antisymmetric
hermitian matrix, satisfies the orbifold boundary condition (BC) given by
\beqn
&&\hskip -1.cm
\begin{pmatrix} A_\mu \cr A_y\end{pmatrix} (x, y_j -y) 
= P_j \begin{pmatrix} A_\mu\cr - A_y\end{pmatrix} (x, y_j + y) P_j^{-1}, ~~
(y_0, y_1) = (0, L), \cr
\noalign{\kern 10pt}
&&\hskip -1.cm
P_0 = \diag (I_{10}, - I_1) ,~~ P_1 = \diag (I_4, - I_7) .
\label{orbifoldBC1}
\eeqn
$P_0$ and $P_1$ break $SO(11)$ to $SO(10)$ and $SO(4) \times SO(7)$, respectively.
In all, the symmetry is broken to $SO(4) \times SO(6) \simeq SU(2)_L \times SU(2)_R \times SU(4)$.  
Write $A_M = 2^{-1/2} \sum_{1 \le j<k \le 11} A_M^{(jk)} \, T_{jk}$
where $SO(11)$ generators satisfy 
$[T_{ij} , T_{kl} ] = i (\delta_{ik} T_{jl}  - \delta_{il} T_{jk} + \delta_{jl} T_{ik} - \delta_{jk} T_{il} )$.
Zero modes of $A_y$ exist only for $A_y^{(j, 11)}$ ($j= 1\sim 4$), which become an $SO(4)$ vector or
$SU(2)_L$ doublet Higgs field in four dimensions.

Fermions are introduced in the bulk in {\bf 32} and {\bf 11} of $SO(11)$, $\Psi_{\bf 32}$ and 
$\Psi_{\bf 11}$. 
We introduce a scalar field in {\bf 16} of $SO(10)$, $\Phi_{\bf 16} $, on the Planck brane.
To make matter content in $\Psi_{\bf 32}$ and $\Phi_{\bf 16} $ transparent, let us adopt
the following representation of $SO(11)$ Clifford algebra 
$\{ \Gamma_j , \Gamma_k \} = 2 \delta_{jk} \, I_{32}$ ($j,k = 1 \sim 11$);
\beqn
&&\hskip -1.cm
\Gamma_{1,2,3} = \sigma^{1,2,3} \otimes \sigma^1 \otimes \sigma^1 \otimes \sigma^1 \otimes \sigma^1, \cr
&&\hskip -1.cm
\Gamma_{4,5} = \sigma^0 \otimes \sigma^{2,3} \otimes \sigma^1 \otimes \sigma^1 \otimes \sigma^1, \cr
&&\hskip -1.cm
\Gamma_{6,7} = \sigma^0 \otimes \sigma^0 \otimes \sigma^{2,3} \otimes \sigma^1 \otimes \sigma^1, \cr
&&\hskip -1.cm
\Gamma_{8,9} = \sigma^0 \otimes \sigma^0 \otimes \sigma^0 \otimes \sigma^{2,3} \otimes \sigma^1, \cr
&&\hskip -1.cm
\Gamma_{10,11} = \sigma^0 \otimes \sigma^0 \otimes \sigma^0 \otimes \sigma^0 \otimes \sigma^{2,3} .
\label{Clifford1}
\eeqn
Here $\sigma^0 = I_2$ and $\sigma^{1,2,3}$ are Pauli matrices.  
Note that $\Gamma_{11} = -i \Gamma_1 \cdots \Gamma_{10}$.
The $SO(11)$ generators in the spinorial representation are given by
$T_{jk}^{\rm sp} = - \onehalf i  \Gamma_j \Gamma_k$.  
In this representation the upper and lower half components of $\Psi_{\bf 32}$ 
correspond to ${\bf 16}$ and $\overline{\bf 16}$ of $SO(10)$.
$\Psi_{\bf 32}$ and $\Psi_{\bf 11}$ satisfy 
\beqn
&&\hskip -1.cm
\Psi_{\bf 32} (x, y_j -y) = 
-P_j^{\rm sp} \, \gamma^5 \, \Psi_{\bf 32} (x,y_j+y) ~, \cr
\noalign{\kern 5pt}
&&\hskip -1.cm
\Psi_{\bf 11} (x, y_j -y) = 
\eta^{\bf 11}_j  P_j \, \gamma^5 \, \Psi_{\bf 11} (x,y_j+y) ~, \cr
\noalign{\kern 5pt}
&&\hskip -1.cm
P_0^{\rm sp} = \Gamma_{11} ~, ~~
P_1^{\rm sp} = - \Gamma_1 \Gamma_2 \Gamma_3 \Gamma_4 
= I_2 \otimes \sigma^3 \otimes I_8 ~.
\label{orbifoldBC2}
\eeqn
Here $\gamma^5= \pm1$ correspond to right- and left-handed Lorentz spinors, and
$\eta^{\bf 11}_j  = \pm 1$.
The action in the bulk is given by
\beqn
&&\hskip -1.cm
S^{\rm bulk}  = \int d^5 x \sqrt{- \det G \, } ~ \Big\{ - \frac{1}{4} \Tr F_{MN} F^{MN} 
+ \cL_{g.f.} + \cL_{gh}  \cr
\noalign{\kern 10pt}
&&\hskip 2.0cm
+ \overline{\Psi}_{\bf 32} \cD (c_{\bf 32}) \Psi_{\bf 32}
+ \overline{\Psi}_{\bf 11} \cD (c_{\bf 11}) \Psi_{\bf 11} \Big\} 
\label{action1}
\eeqn
where $\cL_{g.f.}$ and $\cL_{gh}$ are gauge fixing and ghost terms.
Here $F_{MN} = \dd_M A_N - \dd_N A_M - ig [A_M,  A_N]$, 
$\cD (c) = \gamma^A {e_A}^M D_M - c \sigma '(y)$ and
$D_M = \dd_M + \frac{1}{8} \omega_{MBC} [\gamma^B, \gamma^C] - i g A_M$.

The action for  $\Phi_{\bf 16}$ is given by
\beeq
S_{\Phi_{\bf 16}}^{\rm brane}=
\int d^5x \sqrt{-\mbox{det}G \,} \, \delta(y)
\Big\{ -(D_\mu \Phi_{\bf 16})^\dag D^\mu \Phi_{\bf 16}
-\lambda_{\Phi_{\bf 16}} (\Phi_{\bf 16}^\dag \Phi_{\bf 16}-w^2)^2 \Big\}, 
\label{action2}
\eneq
where $D_\mu \Phi_{\bf 16}= ( \dd_\mu - ig A_\mu^{SO(10)} ) \Phi_{\bf 16} $ and 
$A_\mu^{SO(10)} = 2^{-1/2} \sum_{1 \le j < k \le 10} A_\mu^{(jk)} T_{jk}^{\rm sp} $.
$\Phi_{\bf 16}$ develops VEV.  Without loss of generality we suppose that
the 12th component of $\Phi_{\bf 16}$ develops VEV, $\la \Phi_{\bf 16}^{12} \ra = w \not= 0$,
which 
reduces $SO(4) \times SO(6)$ to $SU(2)_L \times SU(3)_C \times U(1)_Y$.
Generators of $SU(2)_L$ and $SU(2)_R$ are
$T^a_{L/R} = \onehalf( \onehalf \ep^{abc} T_{bc} \pm T_{4 a})$.  In the spinorial representation
\beeq
\Big[ T^a_L , T^a_R \Big] =  \onehalf \sigma^a \otimes 
\bigg[ \begin{pmatrix} 1\cr &0 \end{pmatrix} , \begin{pmatrix} 0\cr &1 \end{pmatrix}\bigg] \otimes I_8 ~, \label{generator1}
\eneq
$( \Phi_{\bf 16}^{11} ,  \Phi_{\bf 16}^{12})$ is $({\bf 1}, {\bf 2})$ of $SU(2)_L \times SU(2)_R$.

One finds later that the zero mode of $A_y$ (4D Higgs field) develops nonvanishing VEV
by the Hosotani mechanism, and $SU(2)_L \times U(1)_Y$ breaks down to $U(1)_\EM$.  
Without loss of generality we suppose that $\la A_y^{4,11} \ra \not= 0$.
The $U(1)_\EM$ charge in the unit of $e$ is given by
\beeq
Q_\EM = T_{12} - \onethird (T_{56} + T_{78} + T_{9,10}) ~.
\label{U1charge}
\eneq
The content of $\Psi_{\bf 32}$ is easily determined by examining $Q_\EM$ in the representation
\eqref{Clifford1} with BC \eqref{orbifoldBC2}.  
The result is summarized in Table \ref{quarklepton}.  
BC at $y=0$ with  $P_0^{\rm sp}$ admits parity-even left-handed (right-handed) modes  only 
for ${\bf 16}$  ($\overline{\bf 16}$) of $SO(10)$, whereas BC  at $y=L$ with 
$P_1^{\rm sp}$ admits parity-even left-handed  (right-handed) modes  only for 
$SU(2)_L$ ($SU(2)_R$) doublets.
In Table \ref{quarklepton} a field with hat has an opposite charge to the corresponding one without hat.
For instance, $u_j$ and $\hat u_j$ have $Q_\EM = + \frac{2}{3}$ and $- \frac{2}{3}$, respectively.
Notice that all leptons and quarks in SM, but nothing additional, appear as zero modes
in $\Psi_{\bf 32}$.  In the $SU(5)$  GUT in 4D, the $\overline{\bf 5}$ ({\bf 10})
multiplet contains $\ell_L$ and $d^c_L$ ($q_L, u^c_L, e^c_L$) so that gauge interactions
alter quark/lepton number and $u_L \go u^c_L$, $d^c_L \go e_L$ etc. transitions are induced.  
In the present case these processes do not occur and proton decay is suppressed. 
Indeed proton decay is forbidden to all order, provided that the $\Psi_{\bf 32}, \Psi_{\bf 11}$ 
fermion number $N_\Psi$ is conserved.  All of $u, d, e^-$ have $N_\Psi=+1$ in the current model. 
Although the fermion number current has anomaly, its effect on proton decay is expected to be
negligible as in the case of baryon number non-conservation in SM.

\begin{table}[h,t,b]
{\small
\begin{center}
\caption{The fermion content of $\Psi_{\bf 32}$ in the representation \eqref{Clifford1} of
$\Gamma_j$ matrices.  
Each $SU(2)_L$ or $SU(2)_R$ doublet, 
from  top to  bottom in  $\Psi_{\bf 32}$, is listed from left to right in the table.
The field with hat has an opposite electric charge to the corresponding field without hat.
Zero modes resulting from the BC in \eqref{orbifoldBC2}  are shown. 
$SO(10)$ and $SU(5)$ content of each field is also indicated.
}
\label{quarklepton}
\renewcommand{\arraystretch}{1.2}
\begin{tabular}{|c|cccccccc|cccccccc|}\hline
Name    
&$\begin{matrix} \nu \cr e \end{matrix}$ &$\begin{matrix} \hat d_1 \cr \hat u_1 \end{matrix}$
&$\begin{matrix} u_3 \cr d_3 \end{matrix}$ &$\begin{matrix} \hat d_2 \cr \hat u_2 \end{matrix}$
&$\begin{matrix} u_1 \cr d_1 \end{matrix}$ &$\begin{matrix} \hat e \cr \hat \nu \end{matrix}$
&$\begin{matrix} u_2 \cr d_2 \end{matrix}$ &$\begin{matrix} \hat d_3 \cr \hat u_3 \end{matrix}$
&$\begin{matrix} \hat d_3' \cr \hat u_3' \end{matrix}$ &$\begin{matrix} u_2' \cr d_2' \end{matrix}$
&$\begin{matrix} \hat e' \cr \hat \nu' \end{matrix}$ &$\begin{matrix} u_1' \cr d_1' \end{matrix}$
&$\begin{matrix} \hat d_2' \cr \hat u_2' \end{matrix}$ &$\begin{matrix} u_3' \cr d_3' \end{matrix} $
&$\begin{matrix} \hat d_1' \cr \hat u_1' \end{matrix}$ &$\begin{matrix} \nu' \cr e' \end{matrix}$
\\
\hline
$\begin{matrix}{\rm Zero}\cr {\rm mode} \end{matrix}$
&$\begin{matrix} \nu_L \cr e_L \end{matrix}$& 
&$\begin{matrix} u_{3L} \cr d_{3L} \end{matrix}$ &~
&$\begin{matrix} u_{1L} \cr d_{1L} \end{matrix}$ &~
&$\begin{matrix} u_{2L} \cr d_{2L} \end{matrix}$ &
& &$\begin{matrix} u_{2R} \cr d_{2R} \end{matrix}$
& &$\begin{matrix} u_{1R} \cr d_{1R} \end{matrix}$
& &$\begin{matrix} u_{3R} \cr d_{3R} \end{matrix}$
& &$\begin{matrix} \nu_R \cr e_R \end{matrix}$
\\
\hline
$SO(10)$
&\multicolumn{8}{c|}{{\bf 16}}  &\multicolumn{8}{c|}{$\overline{\bf 16}$} 
\\
\hline
$SU(5)$
&$\begin{matrix}  \overline{\bf 5} \cr  \overline{\bf 5} \end{matrix}$  
&$\begin{matrix} \overline{\bf 5}  \cr {\bf 10}  \end{matrix}$ 
&$\begin{matrix}  {\bf 10}  \cr {\bf 10}   \end{matrix}$  
&$\begin{matrix}  \overline{\bf 5} \cr  {\bf 10}  \end{matrix}$ 
&$\begin{matrix}  {\bf 10}  \cr {\bf 10}  \end{matrix}$  
&$\begin{matrix}   {\bf 10} \cr {\bf 1}   \end{matrix}$ 
&$\begin{matrix}   {\bf 10}  \cr {\bf 10}   \end{matrix}$  
&$\begin{matrix}   \overline{\bf 5}  \cr {\bf 10}  \end{matrix}$ 
&$\begin{matrix}  \overline{\bf 10} \cr  \overline{\bf 10}  \end{matrix}$  
&$\begin{matrix}  \overline{\bf 10} \cr   {\bf 5}  \end{matrix}$ 
&$\begin{matrix}  {\bf 5}  \cr  {\bf 5}  \end{matrix}$  
&$\begin{matrix}  \overline{\bf 10} \cr   {\bf 5}   \end{matrix}$ 
&$\begin{matrix} \overline{\bf 10} \cr  \overline{\bf 10}  \end{matrix}$  
&$\begin{matrix}  \overline{\bf 10} \cr   {\bf 5}  \end{matrix}$ 
&$\begin{matrix}  \overline{\bf 10} \cr  \overline{\bf 10}  \end{matrix}$  
&$\begin{matrix}  {\bf 1}  \cr  \overline{\bf 10}  \end{matrix}$ 
\\
\hline
\end{tabular}
\end{center}
}
\end{table}

In the gauge field sector,  BC \eqref{orbifoldBC1} alone leads to zero modes (4D massless 
gauge fields)  in  $SO(4) \times SO(6)$, some of which become massive due to 
$\la \Phi_{\bf 16} \ra \not= 0$, leaving only $SU(2)_L \times SU(3)_C \times U(1)_Y$ invariance.
Indeed, $- g^2 | A_\mu^{SO(10)} \la \Phi_{\bf 16} \ra |^2$ 
on the Planck brane generates mass terms of the form 
$\cL_{\rm mass}^{\rm gauge} = - \delta(y) \frac{1}{4} g^2 w^2 (A_\mu^\alpha)^2$.
We assume that $g w / \sqrt{L}$ is much larger than the KK mass 
scale $m_\KK = \pi k z_L^{-1}$.
In this case, even if $A_\mu^\alpha$ is parity even at $y=0$, the boundary condition 
becomes effectively Dirichlet  condition and the lowest KK mode acquires a mass of 
$O(m_\KK)$.\cite{HOOS2008, HNU2010}
It is straightforward to check that all gauge fields in 
$SO(4) \times SO(6) / SU(2)_L \times SU(3)_C \times U(1)_Y$ become massive.
In particular,  among $SU(5)$ diagonal and $SU(3)_C$ neutral components 
\beqn
&&\hskip -1.cm
A_\mu^{3_L} =  \frac{1}{\sqrt{2}} (A_\mu^{12} + A_\mu^{34})  ~, \cr
\noalign{\kern 10pt}
&&\hskip -1.cm
B_\mu^Y = \sqrt{ \frac{3}{10}} (A_\mu^{12} - A_\mu^{34})  - 
 \sqrt{ \frac{2}{15}} (A_\mu^{56} + A_\mu^{78} + A_\mu^{9,10} ) ~, \cr
\noalign{\kern 10pt}
&&\hskip -1.cm
C_\mu =  \sqrt{ \frac{1}{5}} (A_\mu^{12} - A_\mu^{34} 
+  A_\mu^{56} + A_\mu^{78} + A_\mu^{9,10} ) ~,
\eeqn
$C_\mu $ becomes massive due to $\la \Phi_{\bf 16} \ra \not= 0$.
$B_\mu^Y$ is a gauge field of $U(1)_Y$.
After EW symmetry breaking by the Hosotani mechanism, $A_\mu^{34}$ mixes
with $A_\mu^{4, 11}$.  The photon  is given by
\beeq
A_\mu^\EM = \frac{\sqrt{3}}{2} A_\mu^{12} 
- \frac{1}{2 \sqrt{3}} (A_\mu^{56} + A_\mu^{78} + A_\mu^{9,10} )~.
\label{photon}
\eneq
In terms of the  $SU(2)_L$ coupling $g_w = g/\sqrt{L}$ in 4D, 
the $U(1)_\EM$ and $U(1)_Y$ couplings are $e = (3/8)^{1/2} g_w$ and
$g'_Y =(3/5)^{1/2} g_w$.  The Weinberg angle is given by
$\sin^2 \theta_W = 3/8$.

$\la \Phi_{\bf 16} \ra \not= 0$ breaks $SO(10)$ to $SU(5)$ on the Planck brane.
We add a comment that there appear twenty-one would-be NG bosons associated 
with this symmetry breaking,  among which nine of them are eaten by gauge fields in 
$SO(4) \times SO(6) / SU(2)_L \times SU(3)_C \times U(1)_Y$.
There remain twelve uneaten  NG modes corresponding to a complex scalar field 
with the same SM quantum numbers $({\bf 3}, {\bf 2})_{1/6}$   as a quark doublet.
They are massless at the tree level, but would acquire masses at the quantum level.
Further they are color-confined.  It is expected that these colored scalars and quarks 
form color-singlet bound states, whose dynamics can be explored by collider experiments.
The evaluation of masses of these new  bound states, as well as deriving  their experimental 
consequences, is reserved  for future investigation.
We note that  $\la \Phi_{\bf 16} \ra \not= 0$ also gives large brane mass terms for gauge fields
in $SO(10)/SU(5)$, which effectively alters the Neumann BC  at $y=0$ to the Dirichlet BC 
for  their low-lying modes ($m_n \ll gw/\sqrt{L}$).

The extra-dimensional component of gauge fields, $A_y^{a,11}$ $(a=1, \cdots, 4)$, 
admits a zero mode, and  yields  a nonvanishing Aharonov-Bohm (AB) phase playing a role 
of 4D Higgs fields.  AB phases are defined as  phases of  eigenvalues of 
$\hat W = P \exp \big\{ ig \int_{-L}^L dy \, A_y \big\} \cdot P_1 P_0$, which are
invariant under gauge transformations preserving the orbifold BC.\cite{YH1, HHHK2003}
We expand $A_y^{4,11} (x,y)$ as
\beeq
A^{4,11}_y (x,y)=\big\{ \theta_H f_H + H(x) \big\} \,   u_H(y) +\cdots ~, 
\label{AB1}
\eneq
where $f_H = (2/g) \sqrt{k/(z_L^2 -1)}$, 
$u_H(y) = \sqrt{ 2k/(z_L^2 -1)}\,  e^{2ky}$ ($0 \le y \le L$), and $u_H(-y) = u_H(y) = u_H(y+2L)$.
$H(x)$ is identified with the neutral Higgs boson in four dimensions.
Insertion of \eqref{AB1} into $\hat W$ shows that $\theta_H$ is the AB phase.
A gauge transformation generated by
\beeq
\Omega(y; \beta ) 
= \exp \Big\{ -i \beta \,  \frac{z_L^2 - e^{2ky}}{z_L^2 -1} \, T_{4,11} \Big\} ,
\label{gaugeT1}
\eneq
shifts $\theta_H$ to $\theta_H + \beta$, and changes BC matrices  to
$P_0' = e^{-2i \beta T_{4,11}} P_0$ and $P_1'=P_1$.
Note that $T_{4,11} = \sigma^2$ in the 4-11 subspace in the vectorial representation, and 
$T_{4,11}^{\rm sp} = - \onehalf \sigma^0 \otimes \sigma^2 \otimes \sigma^1 \otimes 
\sigma^1 \otimes \sigma^2$ in the spinorial representation. 
The boundary conditions in \eqref{orbifoldBC1} and \eqref{orbifoldBC2} are preserved 
provided $\beta = 2 \pi n$ ($n$: an integer).
The gauge invariance guarantees the periodicity in $\theta_H$ for physical quantities.

The value of $\theta_H$ is determined by the location of the global minimum of 
the effective potential $V_\eff (\theta_H)$, which is flat at the tree level but becomes nontrivial
at the one loop level. 
To find the mass spectra for $\theta_H \not= 0$ and evaluate $V_\eff (\theta_H)$, 
it is most convenient to move to the twisted gauge generated by $\Omega(y; - \theta_H)$.
In this gauge the background $\tilde A_y$ vanishes and $\tilde \theta_H = 0$.
(Quantities with tilde denote those in the twisted gauge.)
The boundary condition matrices become
\beeq
\tilde P_0 =\begin{pmatrix} \cos 2 \theta_H & -\sin 2\theta_H \cr 
       - \sin 2\theta_H & - \cos 2 \theta_H  \end{pmatrix}
\label{BC3}
\eneq
in the 4-11 subspace, and $\tilde P_1 = P_1$.  For $\tilde \Psi_{\bf 32}$
\beeq
\tilde P_0^{\rm sp} =\begin{pmatrix} \cos  \theta_H & -i \sin \theta_H \cr 
       i \sin  \theta_H & - \cos  \theta_H  \end{pmatrix} 
\label{BC4}
\eneq
for pairs $(\nu, \nu')$, $(e,e')$, $(u_j, u_j')$,  $(d_j, d_j')$, whereas 
for pairs $(\hat \nu, \hat \nu')$, $(\hat e,\hat e')$, $(\hat u_j, \hat u_j')$,   $(\hat d_j, \hat d_j')$
$\theta_H \go - \theta_H$ in \eqref{BC4}.   $\tilde P_1^{\rm sp} = P_1^{\rm sp}$.

$\tilde \theta_H=0$ in the twisted gauge  so that all fields satisfy free equations in the bulk
to the leading order  and obey the original boundary conditions at $y=L$.  It is convenient to analyze in the 
conformal coordinate $z \equiv e^{ky}$ ($1 \le z \le z_L$).  Mode functions are expressed
in terms of Bessel functions.  Base functions are tabulated in Appendix A of Ref.\ \cite{LHCsignalsDM}.
For instance, $C(z;\lambda) = \onehalf \pi \lambda z z_L F_{1,0} (\lambda z, \lambda z_L)$
and $S(z;\lambda) = -\onehalf \pi \lambda z  F_{1,1} (\lambda z, \lambda z_L)$
where $F_{\alpha,\beta} (u,v) = J_\alpha (u) Y_\beta (v) - Y_\alpha (u) J_\beta (v)$.

Only particle spectra depending on $\theta_H$ affect the $\theta_H$-dependent part  of $V_\eff$
at 1-loop.  In the gauge field sector, 
$\tilde A_\mu^{a_L}, \tilde A_\mu^{a_R}$ and $\tilde A_\mu^{a, 11}$ $(a=1,2)$ mix with each other.
Their  mass spectra ($m_n = k \lambda_n$)  are determined by zeros of
$C ( 2 S C'+ \lambda \sin^2 \theta_H ) |_{z=1} = 0$ where $C' = dC/dz$.
\beqn
&&\hskip -1.cm
W ~\hbox{tower:} \quad  2 S(1;\lambda_n) C'(1; \lambda_n) + \lambda_n \sin^2 \theta_H = 0 ~, \cr
&&\hskip -1.cm
W_R ~\hbox{tower:} \quad  C(1; \lambda_n) = 0 ~.
\label{Wspectrum}
\eeqn
Similarly
$\tilde A_\mu^{3_L},  \tilde B_\mu^Y , \tilde C_\mu$ and $ \tilde A_\mu^{3, 11}$ mix with each other
whose spectra are given by 
\beqn
&\gamma ~\hbox{tower:} &C' (1; \lambda_n) = 0 ~, \cr
&Z ~\hbox{tower:} & 5 S(1;\lambda_n) C'(1; \lambda_n) + 4 \lambda_n \sin^2 \theta_H = 0 ~, \cr
&Z_R ~\hbox{tower:} & C(1; \lambda_n) = 0 ~.
\label{Zspectrum}
\eeqn
The $Y$ boson part, $\tilde A_\mu^{a j}$ ($a=3,4, 11$, $j=5 \sim 10$), also yields $\theta_H$-dependent
spectra.  It decomposes into 6 sets of $\{ (a,j) \} =\{ (3,5), (4,6), (11,6) \}$, $\{ (3,6), (4,5), (11,5) \}$, etc.
In each set
\beqn
&Y ~\hbox{tower:} & 2 S(1;\lambda_n) C'(1; \lambda_n) + \lambda_n (1 + \cos^2 \theta_H) = 0 ~, \cr
&\hat Y ~\hbox{tower:} & S(1; \lambda_n) = 0 ~.
\label{Yspectrum}
\eeqn
Other components of $\tilde A_\mu$ have $\theta_H$-independent spectra.
The spectra of $\tilde A_z = (kz)^{-1} \tilde A_y$ are simpler, as $\Phi_{\bf 16}$
does not couple to $\tilde A_z$. 
The spectra of $[\tilde A_z^{a4}, \tilde A_z^{a11} ]$  ($a= 1\sim 3, 5 \sim 10$) are given by
\beeq
S(1;\lambda_n) C'(1; \lambda_n) + \lambda_n  
\begin{pmatrix}\sin^2 \theta_H \cr \cos^2 \theta_H \end{pmatrix} = 0 \quad
\hbox{for~} a = \Big\{ \,  \begin{matrix}  1 \sim 3 ~,    \cr  5 \sim 10 . \end{matrix} 
\label{Sspectrum}
\eneq
Other components of $\tilde A_z$ have $\theta_H$-independent spectra.
It follows from Eqs.\ \eqref{Wspectrum} and \eqref{Zspectrum} that  
$m_W \sim (k/L)^{1/2}  z_L^{-1} | \sin \theta_H |$ and $m_Z \sim m_W/\cos \theta_W$ 
for $z_L \gg 1$. 
Wave functions of $W$ and $Z$ are the same as in the $SO(5) \times U(1)_X$ theory
with $\sin^2 \theta_W = 3/8$.

The spectrum of $\tilde \Psi_{\bf 32}$ is determined by the boundary condition  \eqref{orbifoldBC2}
with $\tilde P_0^{\rm sp}$ in \eqref{BC4}.
The spectrum is given, in the absence of brane interactions discussed below,   by
\beeq
 S_L (1; \lambda_n , c_{\bf 32}) S_R (1; \lambda_n , c_{\bf 32}) 
 +\begin{pmatrix} \sin^2 \onehalf \theta_H \cr
 \noalign{\kern 5pt} \cos^2 \onehalf \theta_H \end{pmatrix} = 0 
\label{SpFspectrum}
\eneq
where the upper component is 
for pairs $(\nu, \nu')$, $(e,e')$, $(u_j, u_j')$,  $(d_j, d_j')$ and the lower component 
for pairs $(\hat \nu, \hat \nu')$, $(\hat e,\hat e')$, $(\hat u_j, \hat u_j')$,   $(\hat d_j, \hat d_j')$.
Here $S_{L/R} (z; \lambda, c) = \mp \onehalf \pi \lambda \sqrt{z z_L} 
F_{c \pm \onehalf, c \pm \onehalf} (\lambda z, \lambda z_L)$.  
For $\tilde \Psi_{\bf 11}$, the 4th and 
11th components mix through $\tilde P_0$ in \eqref{BC3}, and their spectrum is given by
\beeq
 S_L (1; \lambda_n , c_{\bf 11}) S_R (1; \lambda_n , c_{\bf 11}) + 
 \begin{pmatrix} \sin^2  \theta_H \cr \cos^2  \theta_H \end{pmatrix} = 0 
\label{VecFspectrum}
\eneq
for $\eta^{\bf 11}_ 0 \eta^{\bf 11}_ 1 = \pm 1$.
Other components have $\theta_H$-independent spectra.
We note that the spectrum of $\Psi_{\bf 32}$ is periodic in $\theta_H$ with a period $2\pi$,
whereas that of gauge fields and $\Psi_{\bf 11}$ with a period $\pi$.

\begin{figure}[htb]
\centering
\includegraphics[bb=0 0 448 286, width=7cm]{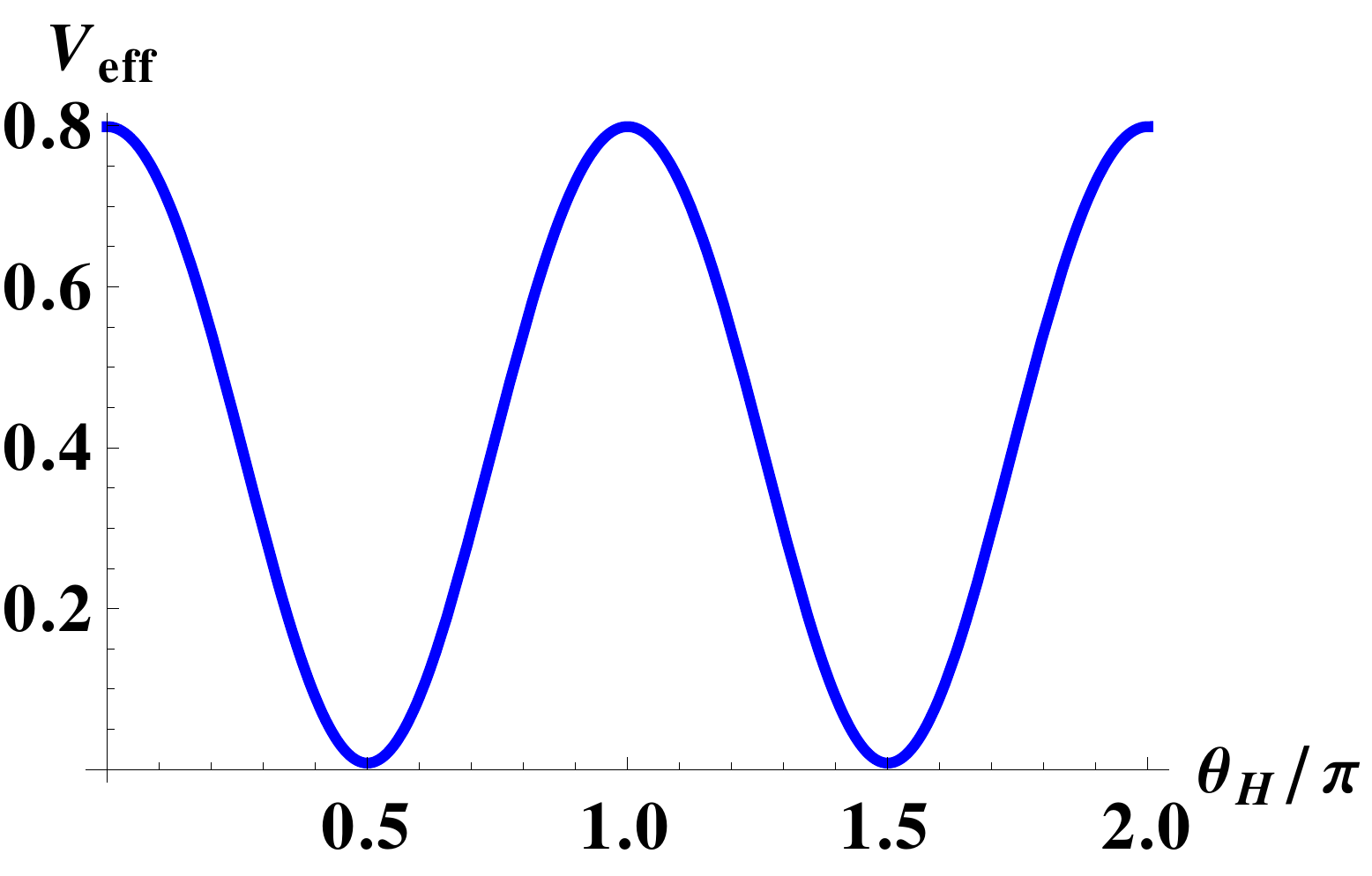}
\caption{$V_\eff (\theta_H)$ in pure gauge theory.  
$U = (4\pi)^2 (kz_L^{-1})^{-4} V_\eff $  is plotted  in the $\xi=0$ gauge. 
The shape of the potential in the $\xi=1$ gauge is almost the same as depicted.
The global minimum is located at $\theta_H = \pm \onehalf \pi$.  
$V_\eff (\theta_H)$ with a minimum at $0< \theta_H < \onehalf \pi$
is achieved with inclusion of fermions and brane interactions.
}
\label{eff-potential1}
\end{figure}

With the mass spectrum at hand, one can evaluate $V_\eff (\theta_H)$ at 1-loop
in the standard method.\cite{HOOS2008,  FHHOS2013}
There is a distinct feature in the spectrum in the gauge field sector.
In the gauge-Higgs grand unification there are six $Y$ towers with the spectrum \eqref{Yspectrum}
where the lowest modes have the smallest mass for $\cos \theta_H=0$.
This leads to an important consequence that even in pure gauge theory the EW symmetry is
spontaneously broken by the Hosotani mechanism.  $V_\eff (\theta_H)$ evaluated with 
\eqref{Wspectrum}-\eqref{Sspectrum} has the global minimum at $\theta_H = \pm \onehalf \pi$.
See Fig.\ \ref{eff-potential1}.
This has never happened in the gauge-Higgs EW unification models.
$\Psi_{\bf 32}$ does not affect this behavior very much in the absence of  brane interactions.
Contributions from particles with the upper spectrum in \eqref{SpFspectrum} and those 
with the lower spectrum almost cancel numerically  in  $V_\eff (\theta_H)$  for $z_L \gg 1$.  
$\Psi_{\bf 11}$ with $\eta^{\bf 11}_0  \eta^{\bf 11}_1 =1$ ($-1$) in \eqref{VecFspectrum} 
strengthens (weakens) the EW symmetry breaking.

At this stage, however, quarks and leptons have degenerate masses.
The degeneracy is lifted by interactions on the Planck brane (at $ y=0$) which must respect
$SO(10)$ invariance.  Let us decompose $\Psi_{\bf 32}$ into {\bf 16} and $\overline{\bf 16}$
of $SO(10)$;  $(\Psi_{\bf 16}^{\rm sp}, \Psi_{\overline{\bf 16}}^{\rm sp})$.
Similarly we decompose $\Psi_{\bf 11}$ into $(\Psi_{\bf 10}^{\rm vec}, \Psi_{\bf 1}^{\rm vec})$.
In terms of these fields with $\Phi_{\bf 16}$,  various $SO(10)$-invariant brane interactions
such as $\overline{\Psi}{}^{\rm sp}_{\bf 16} \Psi_{\bf 1}^{\rm vec} \Phi_{\bf 16}$ and
$\overline{\Psi}{}^{\rm sp}_{\overline{\bf 16}} \Psi^{{\rm vec}}_{\bf 10} \Phi_{\bf 16}$
are allowed on the Planck brane, with which a more realistic fermion spectrum can be achieved.
One may introduce terms like $\overline{\Psi}{}^{\rm vec}_{\bf 1} \Psi^{{\rm vec},c}_{\bf 1}$,
which, in combination of mixing with neutral  components in $\Psi_{\bf 32}$, 
may induce  Majorana masses for neutrinos.   However, it has to be kept in mind that 
such terms may lead to proton decay at higher loops.
As mentioned above, $V_\eff (\theta_H)$ is minimized at $\theta_H= \pm \onehalf \pi$ in pure gauge
theory.  $\theta_H= \pm \onehalf \pi$, however, leads to a stable Higgs boson due to 
the $H$ parity,\cite{HKT2009, HTU2011} which is excluded phenomenologically. 
Desirable value of $\theta_H$ can be achieved by 
appropriate choice of $\eta^{\bf 11}_0  \eta^{\bf 11}_1$ and  inclusion of brane interactions 
for $\Psi_{\bf 32}$ and $\Psi_{\bf 11}$.
Alternatively one may  introduce fermions  $(\Psi_{\bf 55}, \Psi_{\bf 11}, \Psi_{\bf 32})$ 
such that  quarks and leptons are dominantly contained in $(\Psi_{\bf 55}, \Psi_{\bf 11})$.

In this paper we have presented the $SO(11)$ gauge-Higgs grand unification model
which generalizes the $SO(5) \times U(1)_X $ gauge-Higgs EW unification.
The orbifold boundary condition and brane scalar $\Phi_{\bf 16}$ reduce the $SO(11)$ symmetry
directly to the SM symmetry.  The 4D Higgs doublet appears as the extra-dimensional 
component of the gauge potentials with custodial symmetry.  The EW symmetry
is spontaneously broken by the Hosotani mechanism, even in the pure gauge theory.
We presented a model with $\Psi_{\bf 32}$ and $\Psi_{\bf 11}$ for quarks and leptons.
Proton decay is suppressed by the fermion number $N_\Psi$ conservation in the absence
of Majorana masses.  
The effect of the fermion number current anomaly for proton decay is expected to be small.
Although  neutrino Majorana masses lead to proton decay at higher loops, the contribution will
be suppressed by large Majorana masses and loop effect.
There remains a task to pin down the parameters of the model to reproduce
the observed Higgs boson mass and quark-lepton spectrum, and derive phenomenological
predictions.  
Further the masses of the colored would-be NG bosons from $\Phi_{\bf 16}$ 
and color-singlet bound states need to be clarified and the consistency with  experimental
results at LHC need to be examined.
We will come back to these issues  in forthcoming papers.


\subsection*{Acknowledgements}
We thank Nobuhito Maru and the referee for many valuable comments.
This work was supported in part  by  JSPS KAKENHI grants No.\ 23104009 
and No.\ 15K05052.

\vskip .5cm

\ignore{
\renewenvironment{thebibliography}[1]
         {\begin{list}{[$\,$\arabic{enumi}$\,$]}  
         {\usecounter{enumi}\setlength{\parsep}{0pt}
          \setlength{\itemsep}{0pt}  \renewcommand{\baselinestretch}{1.2}
          \settowidth
         {\labelwidth}{#1 ~ ~}\sloppy}}{\end{list}}
}

\def\jnl#1#2#3#4{{#1}{\bf #2},  #3 (#4)}

\def\Zphys{{\em Z.\ Phys.} }
\def\jssc{{\em J.\ Solid State Chem.\ }}
\def\jpsJ{{\em J.\ Phys.\ Soc.\ Japan }}
\def\ptps{{\em Prog.\ Theoret.\ Phys.\ Suppl.\ }}
\def\PTP{{\em Prog.\ Theoret.\ Phys.\  }}
\def\PTEP{{\em Prog.\ Theoret.\ Exp.\  Phys.\  }}
\def\JMP{{\em J. Math.\ Phys.} }
\def\NPB{{\em Nucl.\ Phys.} B}
\def\NP{{\em Nucl.\ Phys.} }
\def\PLB{{\it Phys.\ Lett.} B}
\def\PL{{\em Phys.\ Lett.} }
\def\PRL{\em Phys.\ Rev.\ Lett. }
\def\PRB{{\em Phys.\ Rev.} B}
\def\PRD{{\em Phys.\ Rev.} D}
\def\PRe{{\em Phys.\ Rep.} }
\def\AP{{\em Ann.\ Phys.\ (N.Y.)} }
\def\RMP{{\em Rev.\ Mod.\ Phys.} }
\def\ZPC{{\em Z.\ Phys.} C}
\def\SCI{\em Science}
\def\CMP{\em Comm.\ Math.\ Phys. }
\def\MPLA{{\em Mod.\ Phys.\ Lett.} A}
\def\IJMPA{{\em Int.\ J.\ Mod.\ Phys.} A}
\def\IJMPB{{\em Int.\ J.\ Mod.\ Phys.} B}
\def\EPJC{{\em Eur.\ Phys.\ J.} C}
\def\PR{{\em Phys.\ Rev.} }
\def\JHEP{{\em JHEP} }
\def\JCAP{{\em JCAP} }
\def\cmp{{\em Com.\ Math.\ Phys.}}
\def\JPA{{\em J.\  Phys.} A}
\def\JPG{{\em J.\  Phys.} G}
\def\NJP{{\em New.\ J.\  Phys.} }
\def\CQG{\em Class.\ Quant.\ Grav. }
\def\ATMP{{\em Adv.\ Theoret.\ Math.\ Phys.} }
\def\ibid{{\em ibid.} }

\renewenvironment{thebibliography}[1]
         {\begin{list}{[$\,$\arabic{enumi}$\,$]}  
         {\usecounter{enumi}\setlength{\parsep}{0pt}
          \setlength{\itemsep}{0pt}  \renewcommand{\baselinestretch}{1.0}
          \settowidth
         {\labelwidth}{#1 ~ ~}\sloppy}}{\end{list}}

\def\reftitle#1{{\it ``#1,'' }}    


\end{document}